\documentclass[twocolumn]{aastex63}
\usepackage[utf8]{inputenc}

\usepackage{amsmath}

\newcommand*{\wasyfamily}{\fontencoding{U}\fontfamily{wasy}\selectfont}

\newcommand*{\jupiter}{{\text{\wasyfamily\char88}}}

\begin{document}

\title{The Direct Mid-Infrared Detectability of Habitable-zone Exoplanets Around Nearby Stars}

\author{Zach Werber}
\affiliation{Steward Observatory, University of Arizona, Tucson, AZ}
\author{Kevin Wagner}
\affiliation{Steward Observatory, University of Arizona, Tucson, AZ}
\affiliation{NASA Hubble Fellowship Program $-$ Sagan Fellow}
\author{D\'aniel Apai}
\affiliation{Steward Observatory, University of Arizona, Tucson, AZ}
\affiliation{Lunar and Planetary Laboratory, University of Arizona, Tucson, AZ}

\begin{abstract}
Giant planets within the habitable zones of the closest several stars can currently be imaged with ground-based telescopes. Within the next decade, the Extremely Large Telescopes (ELTs) will begin to image the habitable zones of a greater number of nearby stars with much higher sensitivity$-$potentially imaging exo-Earths around the closest stars. To determine the most promising candidates for observations over the next decade, we establish a theoretical framework for the direct detectability of Earth$-$ to super-Jovian-mass exoplanets in the mid-infrared based on available atmospheric and evolutionary models.
Of the 83 closest BAFGK type stars, we select 37 FGK type stars within 10 pc and 34 BA type stars within 30 pc with reliable age constraints. We prioritize targets based on a parametric model of a planet's effective temperature based on a star’s luminosity, distance, and age, and on the planet's orbital semi-major axis, radius, and albedo. We then predict the most likely planets to be detectable with current 8-meter telescopes and with a 39-m ELT with up to 100 hours of observation per star. Putting this together, we recommend observation times needed for the detection of habitable-zone exoplanets spanning the range from very nearby temperate Earth-sized planets to more distant young giant planets. We then recommend ideal initial targets for current telescopes and the upcoming ELTs.
\end{abstract}


\section{\label{sec:intro}Introduction}

Direct imaging is a powerful technique for the characterization of exoplanets as it provides estimates for many fundamental parameters, such as a planet's effective temperature, radius, and orbital properties. Due to a low transit probability resulting in few transiting planets around nearby stars, direct imaging becomes the most plausible method to study nearby planets in detail. It also is the most plausible method for the detection of an Earth analog. However, these planets are very dim with small separations from their host star, resulting in small planet-to-star contrast ratios requiring telescopes with high sensitivity and angular resolution.

To alleviate this, direct imaging has focused on young ($\lesssim 100$ Myr) giant planets, which are brighter due to retaining a significant amount of heat from formation \citep{youngplanets}. Since the closest young stars are at distances of $\sim$30-100 pc, most studies have also focused on planets whose orbits are wider than $\sim$10 au (e.g. HR 8799 \citealt{hr8799initial,hr8799fourth}; $\beta$ Pictoris \citealt{betapicplanetdiscovery}; HD 95086 \citealt{hd95086}; 51 Eri \citealt{51eri}; and HIP 65426 \citealt{hip65426}).

These studies have focused on imaging in the near-infrared ($\lambda \sim 1-5\mu$m) using the $YJHKL+M$-bands where there is a relatively low level of thermal background noise compared to longer wavelengths. 
The highest ratio of near-infrared planet-to-star contrasts of Sun-like stars is $\sim10^{-10}$ for Earth analogs in the habitable zone \citep{nearmidflux,nearvsmid}. However, few sun-like stars' habitable zones exceed 1\arcsec \; projected separations within the solar neighborhood ($\sim$15 ly). These contrast rations currently cannot be achieved at 1\arcsec~by ground-based 8 meter telescopes.

Instead, the $N^\prime$-band ($\lambda \approx 10$-$12.5 \mu m$) in the mid-infrared presents the most promising wavelength range for near-term prospects of directly imaging low-mass habitable-zone planets. At these longer wavelengths, contrasts for Earth analogs around sun-like stars reach $10^{-7}$ \citep{nearmidflux}, corresponding to the peak of a temperate ($T_{\rm eff}\sim$300K) planet's blackbody spectrum. 

Recent results from the New Earths in the Alpha Cen Region (NEAR; \citealt{nearvsmid}) mission has shown that ground-based observatories with adaptive optics can approach the sensitivity limit for super-Earths in very long ($\sim$100 hr) integrations around the best (i.e., closest) stars. Although such sensitivity can currently only be achieved around a handful of the closest stars, existing observatories can also image the habitable zones of other nearby stars with the ability to detect habitable-zone giant planets. Building on this, the next generation of large, ground-based telescopes such as the Extremely Large Telescope (ELT) with its Mid-Infrared ELT Imager and Spectrograph (METIS), will begin searching for true Earth analogs around the closest stars. 

The ELT is expected to see first light in the late 2020s with METIS as a first-generation instrument. With its 39-meter primary mirror, ELT/METIS will see fainter objects at smaller separations than any of the current 8-meter telescopes (see Section \ref{sec:metis}). For the closest stars, METIS will likely be the first instrument with the sensitivity to directly image temperate, rocky planets. It will also have the ability probe the entire habitable zones of nearby stars, necessitating a framework of theoretical predictions for planet brightness and detectability.

Here, we analyze the 83 closest BAFGK type stars and calculate the predicted $N^\prime$-band contrasts for a range of hypothetical planetary radii. We then examine the predicted contrasts for planets within each star's habitable zone and selected the most promising candidates for direct detection in the $N^\prime$-band. We compare these to sensitivities from NEAR with the Very Large Telescope (VLT) to determine potential targets for deep observations with 8-meter telescopes. Finally, we look ahead to the estimated sensitivities of ELT/METIS to determine the ideal initial targets for observations.




\section{\label{sec:method}Methodology}
\subsection{Stellar Candidates Selection}
We began with a compilation of the Hipparcos (\citealt{hipparcos}), Yale Bright Star (\citealt{yale}), and Gliese (\citealt{gliese}) catalogs\footnote{Database can be found here: \url{https://github.com/astronexus/HYG-Database}}, to get a list of the brightest ($M_V<$8) nearby ($<$30 pc) stars. These catalogs were used since the brightest stars were desired, but are not present in newer catalogs such as GAIA due to saturation. This gave each star's spectral type, distance, location, and the B-V color index. All main sequence stars were selected and divided into groups based on spectral type. FGK type stars were selected if they were within 10 parsecs. BA type stars were selected if they were within 30 parsecs. This extra distance was allowed due to BA stars being typically younger and brighter. Brighter stars were considered better candidates as they correspond to a larger amount of absorbed starlight, making the planets hotter and thus brighter and more detectable. The estimated age and mass ranges, as well as the fluxes of the stars in both the $N^\prime$ and $L^\prime$-bands were then added. If any of these age or mass values were not available from the literature, the star was removed from candidacy. In total, 12 stars were removed, mainly from multi-star systems. For the $N^\prime$-band, data from the Infrared Astronomical Satellite (IRAS; \citealt{iras}) 12 $\mu$m survey or the Wide-field Infrared Survey Explorer (WISE; \citealt{WISE}) W3 ($\sim 12 \mu$m)  was used for the flux. Although the large IRAS beam size presents a risk for contamination, all stars used here have brightness magnitudes of 7 or brighter. For background contamination, a star of similar brightness would have to be very close and would likely have been detected in higher resolution observations, so we take these stellar fluxes to be of just the desired star. For the $L^\prime$-band, the WISE W1 ($\sim 3.35 \mu$m) data was used for the stellar flux.

The temperature, radius, and surface gravity of each star was then recorded based on the listed values in the literature (see Appendix).
This, along with the metallicity values reported in the literature, allowed the spectra of each star to be generated from a model grid, \texttt{BT-Settl} \citep{settl}. In total, 71 stars were selected \textemdash \; 34 BA type stars and 37 FGK type stars. $\alpha$ Cen A (Type G2) was the closest star selected (1.35 pc) while $\iota$ Boo (Type A9) was the furthest (29.8 pc). Temperatures of all stars ranged from $\sim$ 4,000 K to $\sim$ 11,400 K. From the literature, BA stars were considerably younger, having a range of 8 Myr to 972 Myr. FGK stars ranged from 300 Myr to 7.5 Gyr. 

Table \ref{tab:fgktable} lists all FGK type stars used and their properties. Table \ref{tab:batable} lists the stellar properties for all BA type stars used.

\subsection{\label{sec:planets}Planetary Atmospheres and Predicted Contrast Ratios}

We generated contrast predictions for nine planetary masses with the smallest being one Earth mass and the largest being ten Jupiter masses. We assume only two sources of heat for each planet: the stellar flux, which was used to find the planetary equilibrium temperature through radiative equilibrium (Equation \ref{eq:tempeq}), and a net internal source of heat from the planet's interior. We utilize the evolutionary models of \cite{tempcurves}, which we briefly recount here. The models assume that the planets have a core of silicates and iron with an ice layer and (for the giant planets) a H/He envelope. Internal heat was generated based the contraction of both the envelope and the core (again, see \citealt{tempcurves}). These evolutionary tracks were used to provide the effective temperature for a given planet mass with an atmosphere modeled by either \texttt{petitCODE} \citep{petitcode} or \texttt{AMES-Cond} \citep{amescond} and for a given age. These were then used as inputs to generate model planetary spectra. \texttt{AMES-Cond} was used for $5M_\jupiter$ and $10M_\jupiter$ planets while \texttt{petitCODE} was used for less massive planets.

\texttt{AMES-Cond} has been shown to accurately model spectra of planets with effective temperatures below 1300 K and evolutionary calculations at higher effective temperatures \citep{amesvalidate}. However, it is better suited for these higher ($\gtrapprox 2 M_\jupiter$) mass planets used here as it does not cover the evolution of lower mass rocky planets (see \citealt{tempcurves} and Figure 7 therein). \texttt{petitCODE} considers stellar radiation as an additional source of heating making it apt to model low-mass, rocky planets (\citealt{petitvalidate}; Figure 7 in \citealt{tempcurves}). 

Planets were assumed to have solar metallicity and be cloud-free. An albedo of 0.3 was assumed for each planet. Although the true distribution of exoplanet albedos is currently unknown, \cite{albedo} has shown this value to be a reasonable prior based on known Solar System and exoplanet constraints.

The \texttt{petitCODE} models are accurate for planet temperatures down to 150 K \citep{tempcurves}; however, we also use temperatures down to 125 K, noting that temperatures less than 150 K should be used with caution. For older planets that were below this cutoff, the internal temperature was taken to be zero. This internal temperature is then combined with the radiative equilibrium temperature for the total planet's blackbody temperature as:

\begin{subequations}
\begin{equation}
    T_{\rm eq}= T_\star \sqrt{\frac{R_\star}{2a}} (1-A_B)^{0.25}
    \label{eq:tempeq}
\end{equation}
\begin{equation}
    T_{\rm eff}=(T_{\rm eq}^4+T_{\rm int}^4)^{0.25}
    \label{eq:bbtemp}
\end{equation}

\end{subequations}

Where $R_\star$ is the stellar radius, $a$ is the planet's distance to the star, and $A_B$ is the planetary bond albedo.

A blackbody temperature cutoff of 125 K was used as planets this cold would likely not be detectable.
\texttt{AMES-Cond} has a lower reliable temperature of 100 K, but the cutoff of 125 K was used as well. An evolutionary track for one Earth mass was not available, so the internal temperature was set to contribute 10\% of the equilibrium temperature, corresponding to the greenhouse effect (i.e., assuming an Earth-like atmosphere). This was done since the $N^\prime$-band is centered at $\approx 11 \mu m$ and cuts off before the strong CO$_2$ absorption feature at $\lambda>12.8~\mu m$, so we expect to see to the warmer surface. While this is a reasonable approximation (temperature estimates are within 2\% for the Earth-Sun system temperatures), the contrasts predictions produced for this mass should be taken with a larger degree of uncertainty (on the order of the temperature percent uncertainty).

Contrast and brightness predictions were generated for both the $N^\prime$ and $L^\prime$-bands. After determining the planet's blackbody temperature through the combination of radiative equilibrium and internal heat as described above, we then calculated the expected surface brightness of the planet from the model atmospheric spectra. The contrast ($C$) at each separation was then calculated as the ratio of the band-averaged planet's surface brightness ($S_p$) to that of the star ($S$), multiplied by the squared ratio of planet radius-to-star radii ($R$) as:

\begin{equation}
    C=\frac{S_p}{S_\star} \left(\frac{R_p}{R_\star}\right)^2
    \label{eq:contrast}
\end{equation}

Where the subscript $p$ means it is a property of the planet.

The planetary radii were given by the evolution tracks of \cite{tempcurves}. For $1M_\earth$, a constant radius of $1R_\earth$ was used. The surface gravity was then calculated to be self-consistent with the radii for the given mass.

\begin{figure}[ht!]
    \centering
    \epsscale{2}
    \plottwo{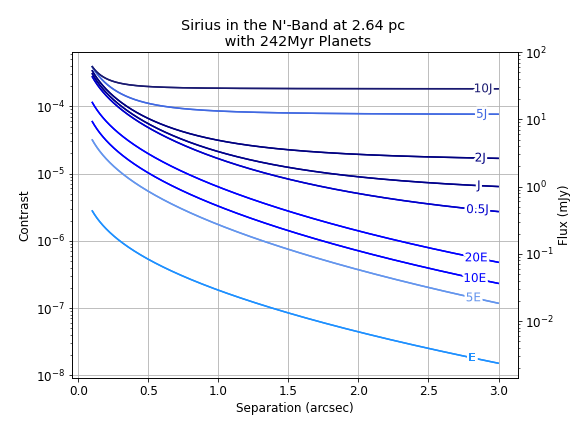}{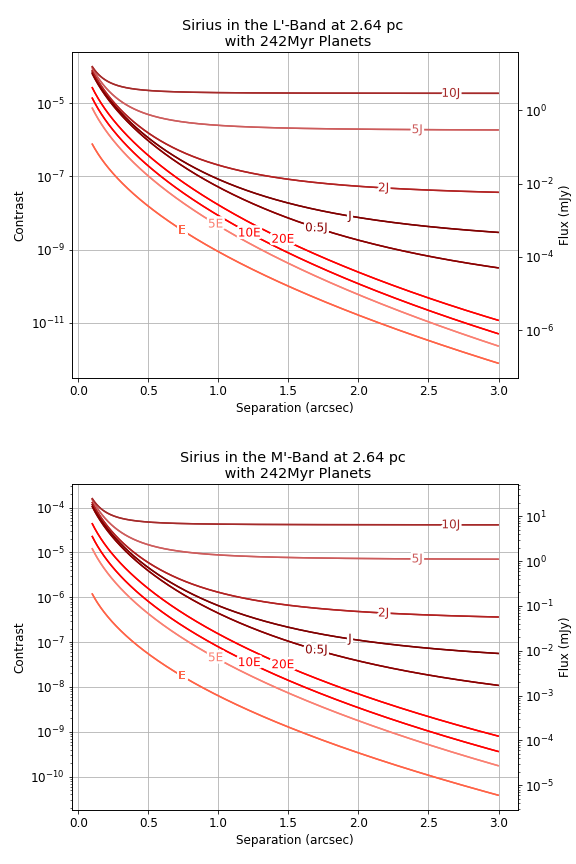}
    \caption{Comparison of contrast ratios for all nine planets in the $N^\prime$-band, $M^\prime$-band $L^\prime$-band. The labels on each line correspond to the mass.}
    \label{fig:compare}
\end{figure}

While most previous observations have been conducted in the $L^\prime$-band, the $N^\prime$-band yields higher planet-to-star contrasts and can more likely lead to direct detection compared to both the $L^\prime$ and $M$-band, as shown in Figure \ref{fig:compare}. Curves were generated from 0.1 out to 3 arcseconds for each star, provided that temperatures were within our limits to that point. Planets that were not above our temperature limit of 125 K were not plotted. For each band, a synthetic box filter was used and the received stellar flux was approximated as a greybody. The box filter was tested for step sizes of 0.05 $\mu$m and 0.1 $\mu$m and the difference was found to be negligible beyond computing time, so a 0.1 $\mu$m filter was used for each calculation. 

\section{\label{sec:results}Results}
For all tested planetary systems, both the contrasts and fluxes were higher in the $N^\prime$-band compared to the $L^\prime$-band, as expected. For simulated Super-Earth mass planets (M~$\lesssim10M_\oplus$), contrasts in the $N^\prime$-band were, on average, $\sim10^{-8}$ at a separation of 1\arcsec \;around BA type stars, while the $L^\prime$-band rarely exceeded a contrast on the order of $10^{-14}$ at the same separation. Only the closest FGK type stars ($\alpha$ Cen A, B, Procyon) had simulated Super-Earths above our cutoff temperature of 125 K (See Section \ref{sec:planets}) at 1\arcsec. However, contrasts around those three stars were significantly higher (see Table \ref{tab:fgkearthfluxes} and Figures \ref{fig:allfluxes}) and \ref{fig:namedfluxes}. 

\begin{figure}[ht!]
    \centering
    \epsscale{1.2}
    \plotone{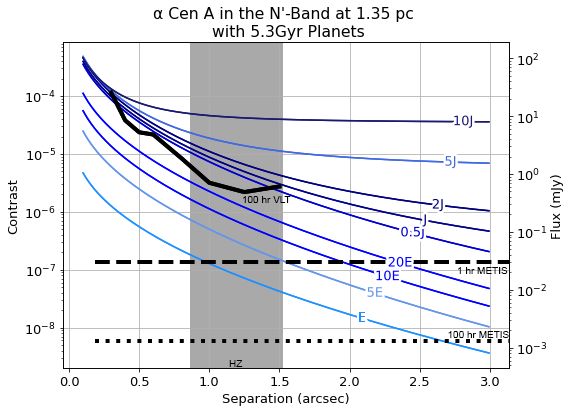}
    \caption{Contrast curves for planets around $\alpha$ Cen A. The ratio of planet-to-star flux (contrast) is shown on the left while the brightness in mJy is shown on the right. The black line represents a SNR $\sim$3 sensitivity curve from \cite{nearvsmid}. The dashed (dotted) black line shows the SNR$\sim$5 flux sensitivity for a 1 hr (100 hr) METIS observation for the background limited region beyond 0.2\arcsec. The habitable zone is shown in grey \citep{hzcalc1}.}
    \label{fig:near}
\end{figure}

The brightest simulated Jupiter-mass planet was around Sirius and could reach a flux of $\sim 3$ mJy at 1\arcsec \; separation (well within the inner habitable zone separation of  $\sim 2\arcsec$) when viewed in the $N^\prime$-band. Based on sensitivity estimates of the Very Large Telescope (VLT) from NEAR \citep{siriusnear} and Large Binocular Telescope (LBT: \citealt{alphacendetect}), these planets would be bright enough to be above the sensitivity cutoff, making this star a promising candidate for direct imaging missions. Our data on $\alpha$ Centauri A (Figure \ref{fig:near}) is in good agreement with \cite{nearvsmid}, which shows that Jupiter mass planets in the habitable zone could be detected (see Figure 2 therein) with 100 hours of observation.

Contrasts were generally consistent within spectral types for similar distance. As expected, younger stars had more bright ($\gtrapprox$ 1 mJy) planets which gave promising contrast ratios for direct imaging. Despite being further, the extra brightness of the modeled Jupiter-mass planets around BA type stars due to their younger ages allowed for the contrast ratios to be similar or higher compared to values for the modeled Jupiter-mass planets around the cooler, closer stars. Younger simulated planets had flatter contrast curves due to the internal temperatures that were nearly equivalent to those created by the stellar flux (see Figure \ref{fig:tempcutoffs}).

Results were tabulated according to mass and spectral type.
Tables \ref{tab:fgkfluxes} and \ref{tab:bafluxes} list the fluxes and contrasts for the simulated one, five, and ten Jupiter mass planets at 1\arcsec \; separations for FGK and BA type stars, respectively. Tables \ref{tab:fgkearthfluxes} and \ref{tab:baearthfluxes} list the fluxes and contrasts for the simulated one, five, and ten Earth mass planets at 1\arcsec \; separations for FGK and BA type stars, respectively.
Tables \ref{tab:fgkhabzone} and \ref{tab:bahabzone} list the fluxes for modeled $1M_\earth$, $10M_\earth$, and $1M_\jupiter$ planets at both the outer and inner edge of the star's habitable zone. Habitable zone distances were calculated based on \citealt{hzcalc1,hzcalc2} for FGK type stars while we used liquid water boundaries for an Earth analog (size and atmosphere) for the habitable zones around BA-stars.

Only simulations of $\alpha$ Cen A and Sirius had $1 M_\jupiter$ planets with fluxes above 1 mJy. However, several FGK and BA type stars had simulated 5 and 10 $M_\jupiter$ fluxes above 1 mJy (reachable currently in a few nights). Stars with these planets are mentioned below as they are currently detectable with the VLT and LBT. We also examine possible habitable zone planet detections with the Mid-Infrared ELT Imager and Spectrograph (METIS) on the Extremely Large Telescope (ELT) as a general case study for the capabilities of future telescopes.

\subsection{BA Star Candidates}
\begin{figure}[ht!]
    \epsscale{1.2}
    \plotone{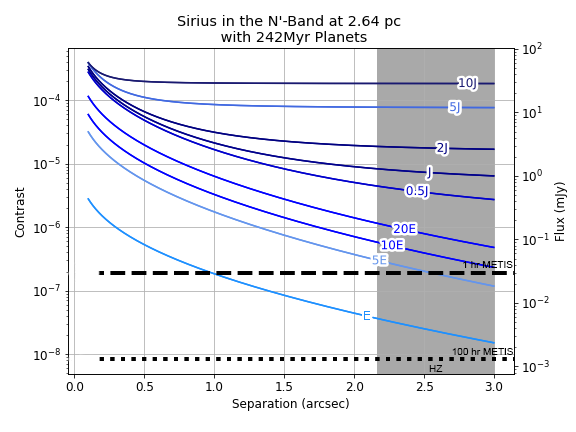}
    \caption{Contrast curves for planets orbiting Sirius in the $N^\prime$-band. The contrast ratio of planet to star is shown on the left and the flux is shown on the right. The habitable zone around Sirius is shown in grey. The dashed (dotted) black line shows the SNR$\sim$5 flux sensitivity for a 1 hr (100 hr) METIS observation for the background limited region beyond 0.2\arcsec.}
    \label{fig:sirius}
\end{figure}

Of the BA type stars, \object{Sirius} (Type A0, 2.6 pc) was both comparatively close and hot, leading to both high contrasts and fluxes ($10^{-5}$ and $\sim$ 3 mJy for a simulated Jupiter mass planet) at a 1\arcsec \; separation (See Figure \ref{fig:sirius}). Similar to $\alpha$ Cen, its proximity leads to smaller planets also being detectable in very long integrations. With comparable sensitivity to NEAR ($\sim$0.3 mJy at 1\arcsec), a $\sim$10-20 M$_\earth$ planet could be imaged with the VLT or LBT.

\object{Vega} (Type A0, 7.7 pc) and \object{Fomalhaut} (Type A3,  7.7 pc) are similar in distance, age, and temperature. Fomalhaut has one potentially imaged planet (\citealt{Fomalhaut,fomalnew}) well outside the star's habitable zone. Vega has been observed extensively since the discovery of its circumstellar disk \citep{vegadisk} and has one close candidate planet \citep{Vega}. Around 1\arcsec, fluxes for a 5 $M_\jupiter$ planet would be 0.77 mJy for Vega and 0.57 mJy around Fomalhaut, which would be detectable with the VLT or LBT in $\sim$25 hours.

\begin{figure}[ht!]
    \centering
    \epsscale{1.2}
    \plotone{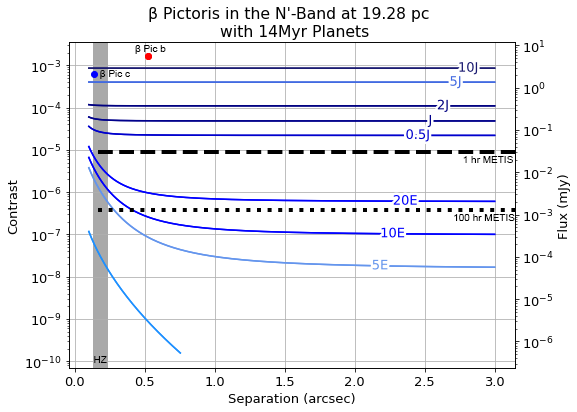}
    \caption{Contrast curves with fluxes for planets orbiting $\beta$ Pic. The habitable zone for $\beta$ Pic is within the grey box and both $\beta$ Pic b \& c have been included as red and blue points, respectively, at their maximum projected separations. Data from \cite{betapicbdata} and \cite{betapiccdata} were used for $\beta$ Pic b and c, repsectively. The dashed (dotted) black line shows the SNR$\sim$5 flux sensitivity for a 1 hr (100 hr) METIS observation for the background limited region beyond 0.2\arcsec.}
    \label{fig:betapiccontrast}
\end{figure}

\object{$\beta$ Pictoris} (Type A6, 19.3 pc) was the youngest star (14 Myr) in our sample. Both a 5 and a 10 $M_\jupiter$ planet would have fluxes above 1 mJy at 1\arcsec. While extensive imaging of the system has been done for both the disk and known giant planets (\citealt{betapicdetection,betapic}), 
current angular resolution limitations prevent imaging of the habitable zone (2.5 to 4.4 AU or 0.13\arcsec\; to 0.23\arcsec), but this separation falls within the inner working angle (IWA) of METIS \citep{metisiwa}. $5-10M_\jupiter$ planets have flat contrast curves due to high internal heating at such a young age (14 Myr). In other words, the known planets should be readily detectable with METIS. Figure \ref{fig:betapiccontrast} shows the contrasts for planets around $\beta$ Pic with its two known planets shown as points at their respective maximum projected separations.

\object{Altair} (Type A7, 5.1 pc) was the only other BA type star to have a simulated $5M_\jupiter$ planet with a flux above 0.7 mJy at a separation of 1\arcsec. Other BA type stars with simulated $10M_\jupiter$ planets with a fluxes above 0.7 mJy at 1\arcsec \; separation were \object{Denebola} (Type A3, 11.1 pc), $\beta$ Ari (Type A5, 18.3 pc) and \object{Alpheratz} (Type B9, 29.8pc). For that separation, those planets would likely be detected within $\sim$25 hours with the VLT and LBT.

Table \ref{tab:bafluxes} contains the complete list of the 34 BA type stars with the contrast and flux of the simulated giant planets at 1\arcsec \; and Table \ref{tab:baearthfluxes} shows the simulated rocky planets and their contrasts and fluxes at the same separation of 1\arcsec \; for completeness.

\subsection{FGK Star Candidates}
FGK type stars were generally older ($\gtrapprox1$ Gyr) and dimmer ($m_V>4$) compared to BA type stars of similar distance. At older ages, planet heat is dominated by stellar flux, meaning dimmer stars result in cooler and fewer detectable planets (see Figure \ref{fig:tempcutoffs}).
The NEAR campaign \citep{nearvsmid} has demonstrated sensitivity to $\sim$30 M$_\earth$ planets in the habitable zone of $\alpha$ Cen A with $\sim$100 hour observations. Figure \ref{fig:near} shows our contrasts compared to the sensitivity of the VLT from NEAR, showing that planets less massive than Jupiter could be observed throughout the habitable zone. We find that at 1\arcsec, an Earth analog would be a factor of $\sim$25 in terms of contrast from the VLT sensitivity.

\object{$\alpha$ Cen B} (Type K1, 1.35 pc) is dimmer than its companion, but still would have detectable planets due to its close proximity. With a modeled flux of 1.3 mJy at 1\arcsec (just outside its habitable zone), a 5$M_\jupiter$ planet would be readily detectable within 100 hours with the VLT ($\alpha$ Cen is too far south to be observed with the LBT). 

\object{$\epsilon$ Eri} (Type K2, 3.2 pc) and \object{Procyon} (Type F5, 3.5 pc) are similar distances, but different spectral types. A Jupiter mass planet around Procyon would be detectable at 1\arcsec despite being being older $-$1.87 Gyr \citep{procyonage} to 600 Myr \citep{agesource}. $5M_\jupiter$ planets around either would be above 0.7 mJy, which current instruments would be able to detect with long ($\sim$25-100 hr) exposures.

\object{70 Oph} (Type K0, 5.1 pc), and \object{$\chi$ Ori} (Type G0, 8.7pc) were the other FGK type stars with $5M_\jupiter$ fluxes above 0.7 mJy at 1\arcsec. \object{$\epsilon$ Indi} (Type K5, 3.6pc), \object{$\xi$ Boo} (Type G8, 6.7pc), and \object{$\pi^3$ Ori} (Type F6, 8.0 pc) all had simulated 10$M_\jupiter$ planets with fluxes above 0.7 mJy at 1\arcsec \; separation, making all of these potential planets currently detectable.

Table \ref{tab:fgkfluxes} contains the complete list of the 37 FGK type stars with a variety of simulated Jupiter mass planets' contrast and flux at 1\arcsec \; and Table \ref{tab:fgkearthfluxes} shows the simulated Earth masses and their contrasts and fluxes at the same separation of 1\arcsec.

\subsection{\label{sec:metis}METIS Sensitivity}
For the $N2$-band ($\lambda \sim 11.2 \mu m$), METIS will have an IWA of 0.07\arcsec \; and will achieve a sensitivity to 30 $\mu$Jy point sources with an SNR of 5 in one hour \citep{metissens}, or 1.3 $\mu$Jy in 100 hours (ESO provided imaging ETC)\footnote{While the actual performance of METIS is not yet known, a factor of ten improvement would be expected by scaling of SNR with exposure time between 1 hr and 100 hr observations. The slightly more optimistic expectation for the 100 hr sensitivity (per the ETC) could be realistic given more substantial field rotation per observation, combined with averaging over independent speckle patterns.} in seeing limited observations at separations of 1\arcsec (\citealt{metisiwa,metissens}). Looking at Tables \ref{tab:fgkhabzone} and \ref{tab:bahabzone}, the habitable zones around all of the BA type stars could be imaged as could those of all of the FGK type stars except Gliese 250 A, Gliese 338 B, and Groombridge 1830.

With a sensitivity of 1.3 $\mu$Jy, any Jupiter mass planet in the habitable zone around all tested stars would be detectable, though some may require multi-night exposures. For the closest stars, such as $\alpha$ Cen A\&B, $\epsilon$ Eri, Procyon, Sirius, Altair, and Fomalhaut, a Jupiter mass planet could be detected within a one hour observation. 
Potential $1M_\jupiter$ planets could be detected around 8 BA-stars and 13 FGK-stars within just one hour each.

\begin{figure}
    \centering
    \epsscale{1.3}
    \plotone{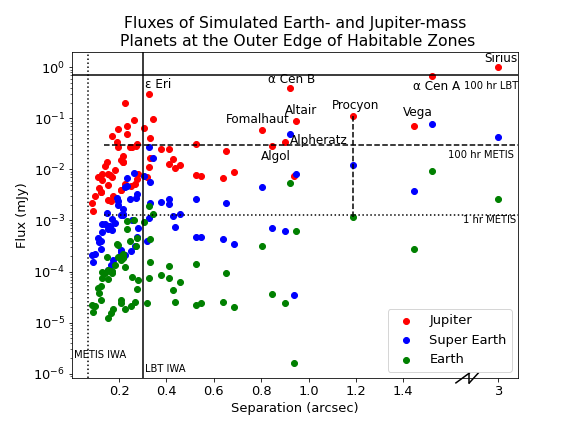}
    \caption{Fluxes of $1M_\jupiter$, $10M_\earth$, and $1M_\earth$ planets at the outer edge of their star's habitable zone. Solid black lines represent current capabilities of the VLT for 100 hour observations and an IWA of $\sim$0.3. The dotted vertical line shows the ELT's expected IWA and the dotted horizontal line is for its expected background limited (separation $\gtrapprox 0.2\arcsec$) sensitivity for an SNR of 5 for 100 hour observations in the $N2$-band. The dashed horizontal line shows the expected background limited (separation $\gtrapprox 0.2\arcsec$) sensitivity for an SNR of 5 for 1 hour observations with METIS with the $N2$-band. Planets around the same star share the same separation and planets for specific stars should be read as vertical lines, with a line connecting planets around Procyon shown. Habitable zone distances were calculated based on \citealt{hzcalc1,hzcalc2} for FGK type stars while we used liquid water boundaries for an Earth analog (size and atmosphere) for the habitable zones around BA type stars.}
    \label{fig:allfluxes}
\end{figure}

\begin{figure}
    \centering
    \epsscale{2.3}
    \plottwo{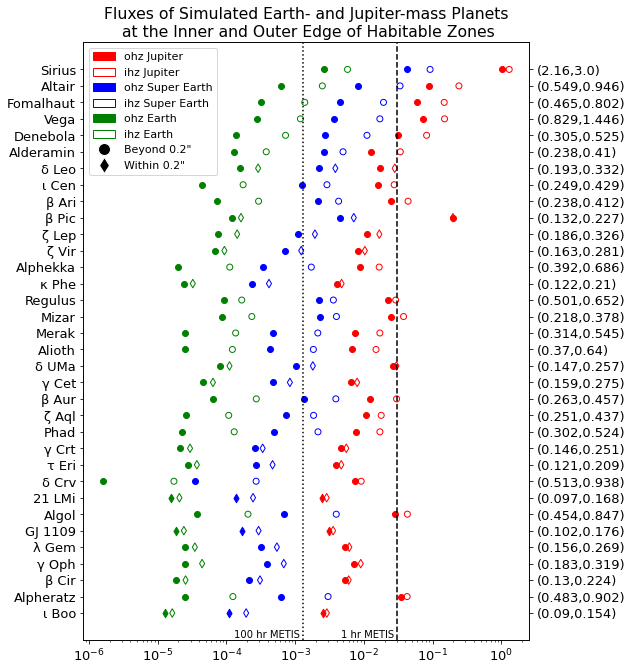}{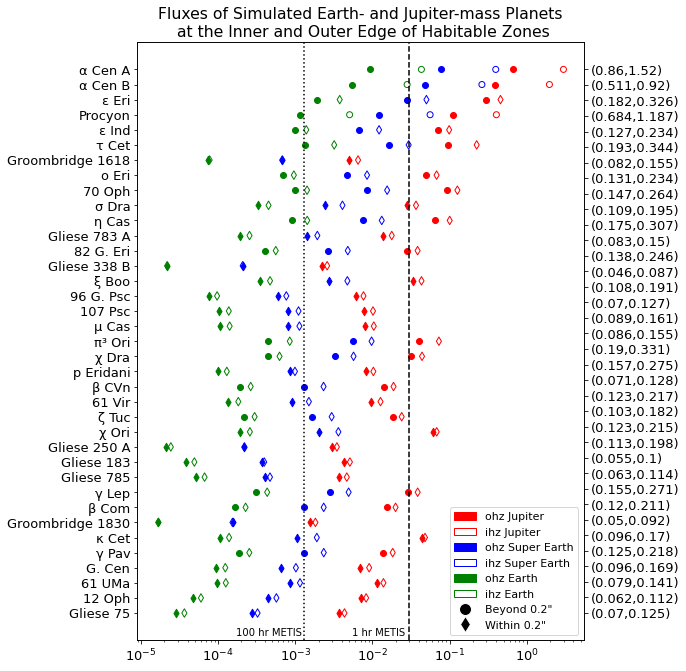}
    \caption{Habitable-zone fluxes of $1M_\jupiter$, $10M_\earth$, and $1M_\earth$ planets for BA type stars (top) and FGK type stars (bottom). Solid shapes represent the flux at the outer edge of the habitable zone (ohz) while the rings represent the inner edge of the habitable zone (ihz). A planet's brightness range throughout the habitable zone can be followed by connecting the open-faced shapes to their like-colored solid shapes. The separation (in arcsec) of the inner and outer edge for each star is shown on the right. The dotted vertical line is for the ELT's expected sensitivity for an SNR of 5 for a 100 hour observation in the $N2$-band for separations greater than 0.2\arcsec. The dashed vertical line shows the expected sensitivity for an SNR of 5 for a 1 hour observation with METIS with the $N2$-band for separations greater than 0.2\arcsec. Planets not within the background limited sensitivity region are plotted as diamonds. Distances to the star increase down the figure. See Tables \ref{tab:fgktable} and \ref{tab:batable} for the distances to each star. Habitable zone distances were calculated based on \citealt{hzcalc1,hzcalc2} for FGK type stars while we used liquid water boundaries for an Earth analog (size and atmosphere) for the habitable zones around BA type stars.}
    \label{fig:namedfluxes}
\end{figure}

\begin{figure*}
    \centering
    
    \plotone{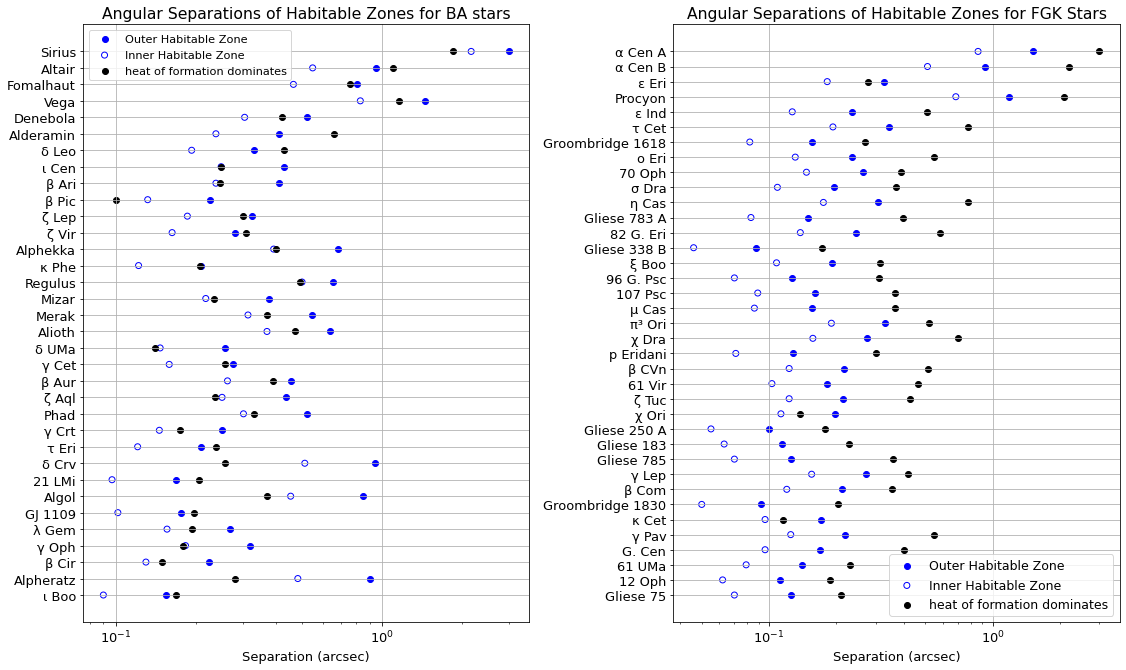}
    \caption{Habitable-zone angular separations for BA type stars (left) and FGK type stars (right). The black circles show the separation at which the remnant heat of formation accounts for greater than 50\% of a $1M_\jupiter$ planet's heat, where all separations to the left of the black circle are dominated by stellar flux. FGK type stars have stellar flux dominate throughout their habitable zones, while BA type stars typically have remnant heat of formation as a dominant heat source for planets through their habitable zones.}
    \label{fig:tempcutoffs}
\end{figure*}

With this sensitivity, METIS will likely begin to image a variety temperate planets around nearby stars. $1M_\earth$ planets at the outer edge of the habitable zone around $\alpha$ Cen A\&B would be detectable within $\sim$3 and 7 hours, respectively. 1 BA type star (Sirius) and 4 FGK type stars would have $1M_\earth$ planets detectable within 100 hours of observations. Figure \ref{fig:allfluxes} shows the fluxes of $1M_\jupiter$, $10M_\earth$, and $1M_\earth$ planets at the outer edge of their star's habitable zone. Horizontal lines for 100 hour VLT and METIS observations as well as 1 hour METIS observations are included. Figure \ref{fig:namedfluxes} shows fluxes of both inner and outer edge habitable zone simulated planets.

\cite{metisiwa} modeled METIS sensitivity around an L=5 magnitude star where the noise is dominated by the diffraction halo (see Figure 13 therein). In this regime, sensitivity was approximately an order of magnitude worse in terms of brightness for very close separations ($\lesssim 0.1\arcsec$) and approximately constant for larger separations ($\gtrsim 0.2\arcsec$). In all figures with METIS sensitivities labeled, these were considered as constant for separations to 0.2\arcsec. Our model began at 0.1\arcsec \; separation and no star discussed below had readily (within 5 hours) detectable simulated planets with separations closer than 0.2\arcsec, even if METIS was background-limited down to 0.1\arcsec.

\cite{metisyield} estimated yields with METIS using 1 hour of integration with its $N2$, $L$, and $M$-bands. As found here as well, their closest stellar candidates would have planets higher contrasts, making them more likely to be detectable. While only the $N2$-band on METIS is explored in this paper, \cite{metisyield} finds this band outperforms the $L$ and $M$-bands and has a 71.1\% chance of making at least one detection with just the 1 hour integration. \cite{metisyield}, however, examines a smaller sample of stars, only testing for A-K type stars within 6.5 parsecs using the High-contrast ELT End-to-end Performance Simulator (HEEPS; \citealt{heeps}) for 1 hour integration. We extend this further out (10-15 pc) with more stars (B-K) and examine longer total observations.

\newpage

\section{\label{sec:discussion}Discussion}

\subsection{BA Stars}
For the 34 BA-type stars we looked at, only Sirius, due to its brightness and proximity, had the possibility for current observations to detect a $1M_\jupiter$ planet within 100 hours  both within the habitable zone and on wider orbits at 1\arcsec. These stars are much hotter and typically younger than FGK-type stars, presenting an interesting prospect from a detection standpoint compared to the more common FGK-type stars. 

The age in particular should increase the planet-to-star contrast ratio due to the additional self-luminosity of the planets, making them more detectable. These stars additionally have their habitable zones at further distances, thus decreasing the necessary angular resolution for detection and again demonstrating the prospective interest for future direct imaging observations. 

The major limitation of observations of most BA stars comes from the available stellar population$-$with only a small number of these stars within the solar neighborhood. Four A type stars are within 10 parsecs and no B type stars are within 10 parsecs. Although we considered BA type stars out to 30 parsecs, this distance, in practice, is better suited for direct imaging of more massive companions further away from their host star. Despite having larger habitable zones, the larger distances to the majority of BA-stars translate to smaller angular separations from the host star$-$with some only extending to  0.2\arcsec, which is not currently resolvable in the $N$-band. However, there are noteworthy exceptions.

Seven BA-stars had hypothetical $10M_\jupiter$ planets that would be detectable at 1\arcsec \; with current telescopes and 4 stars had $5M_\jupiter$ planets that would be detectable at 1\arcsec. Those four possible/hypothetical planets were around the three of the four closest BA type stars (Sirius, Altair, and Vega) and the youngest star ($\beta$ Pic).

Sirius, the closest A-type star, would have the brightest and highest contrast giant planets. Current telescopes also have the capability to image its habitable zone in $N$-band. Note that the effects of Sirius B on the planets were not considered. Based on its relative orbit with Sirius A, Sirius B would contribute less than 1 K to the equilibrium temperature, so we consider its effects to be negligible. Figure \ref{fig:sirius} includes a second plot with 120 Myr planets as the planetary evolutions may have restarted when Sirius B became a white dwarf.

Current telescopes also have the angular resolution to image the habitable zones around Altair, Fomalhaut, and Vega in $N$-band, meaning that possible $5M_\jupiter$ planets around Vega and Altair could currently be detected with Fomalhaut taking slightly longer than 100 hours. Fomalhaut is known to have an excess infrared emission due to a debris disk, indicating that there likely is some planetary system around Fomalhaut based on its age of 536 Myr \citep{fomalhautdisk}. The radiation in the infrared from this disk should be negligible compared to any Jupiter mass planets with internal heat.

Close to Vega's habitable zone at $\sim$14 AU away from the star is a suspected asteroid belt analog \citep{vegabelt}. Observations interior to this possible asteroid belt have revealed a lack of dust, suggesting that a possible planet in the habitable zone could have been the mechanism for clearing this area \citep{vegahzdust}. One candidate planet much closer in than the habitable zone has also been detected through radial velocity measurements \citep{Vega}.

The habitable zone of $\beta$ Pic has a small angular extent from the star, only extending to 0.23\arcsec, thus preventing current $N$-band imaging. Although $\beta$ Pic has a debris disk, the disk surface brightness in the $N$-band will be negligible compared to the Jupiter mass planets with internal heat, similar to the disk around Fomalhaut. The ELTs will be able to resolve  $\beta$ Pic's habitable zone.

The ELTs will also greatly expand the number of stars that planets could potentially be imaged around and significantly lower the mass and separation of planets that can be detected. With the habitable zone around all BA type stars possible to be imaged, all Jupiter mass planets could be imaged and the closest stars could have Earth-sized planets imaged.

\begin{figure*}
    \epsscale{1.17}
    \plottwo{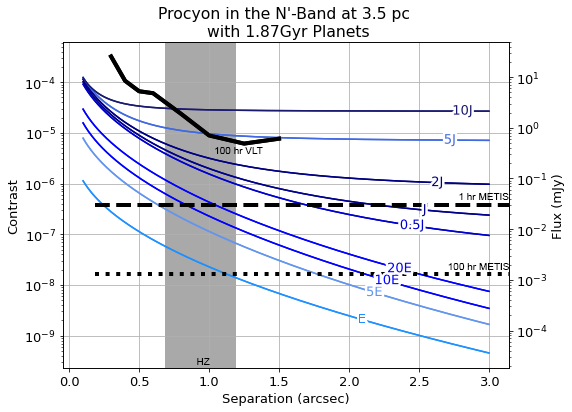}{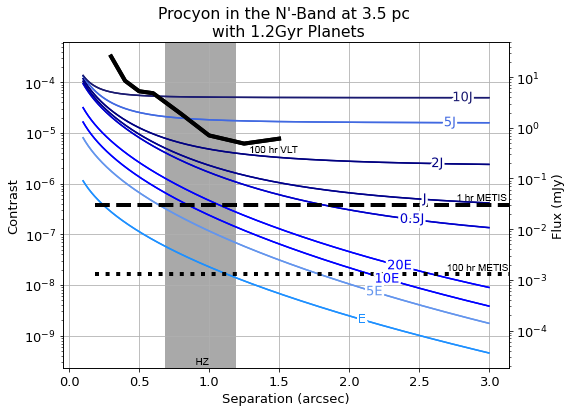}
    \caption{Contrast curves for theoretical planets around Procyon. The black curve shows a SNR of $\sim$3 from \cite{nearvsmid} in terms of brightness. The dashed (dotted) black line shows the SNR$\sim$5 flux sensitivity for a 1 hr (100 hr) METIS observation for the background limited region beyond 0.2\arcsec. The habitable zone is shown in grey. Left: Current age Procyon. Right: Procyon with 1.2 Gyr planets when Procyon B became a white dwarf.}
    \label{fig:procyon_sens}
\end{figure*}

\subsection{FGK Stars}
FGK type stars are typically much closer than BA type stars and are more abundant, but also dimmer. Of the 37 FGK type stars we looked at, only $\alpha$ Cen A had a $1M_\jupiter$ planet bright enough at 1\arcsec \; to be potentially imaged. Since many of these stars are older than 1 Gyr, internal heat from formation is less significant, and therefore planet flux and contrast are primarily distance dependent, making the closer stars far more promising for potential planet detections.

$\alpha$ Cen A and $\alpha$ Cen B have remained popular targets for exoplanet searches due to their proximity to Earth. $\alpha$ Cen A has also been imaged in the mid-infrared with sensitivities extending down to warm Neptune-sized planets throughout the habitable zone \citep{nearvsmid}. Figure \ref{fig:near} shows our contrast predictions in comparison to the NEAR sensitivity measurements and shows good agreement in their findings of sub-Jupiter mass planets being detectable within the habitable zone.

$\epsilon$ Eridani is known to have one Jupiter mass planet at roughly 1\arcsec \; separation \citep{epserib}. Multiple debris belts have been imaged in this system, with the innermost debris belt located just within the orbit of $\epsilon$ Eri b at $\sim$3 AU \citep{epseridisk}. The habitable zone would be within the innermost debris belt at 0.58 to 1.0 AU (0.18\arcsec\;to 0.32\arcsec), leaving most of the habitable zone currently unable to be resolved. 5 and 10 $M_\jupiter$ would be above 1 mJy in brightness, meaning that METIS would be able to easily detect planets. Using the parameters given by \cite{newepseri} for $\epsilon$ Eridani b, the planet would have a separation beyond 1\arcsec \; and expected to be detectable by the $JWST$ (see \cite{epseriimaging}) or within one hour of observations with METIS. 

\begin{figure*}
    \epsscale{1.17}
    \plottwo{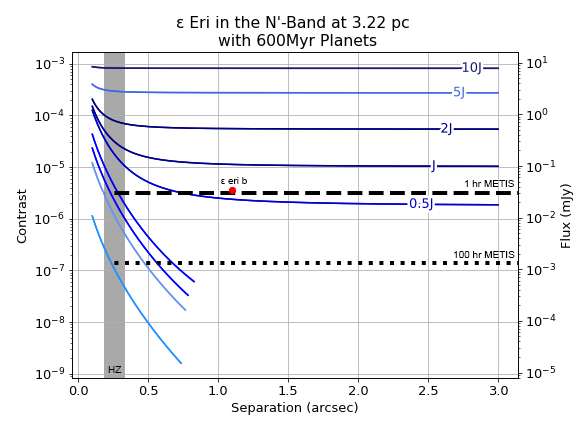}{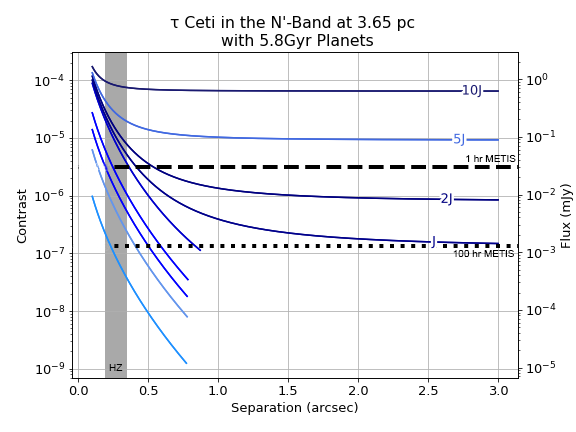}
    \caption{Contrast curves for theoretical planets around $\epsilon$ Eridani (left) and $\tau$ Ceti (right). The dashed (dotted) black line shows the SNR$\sim$5 flux sensitivity for a 1 hr (100 hr) METIS observation for the background limited region beyond 0.2\arcsec. The habitable zone is shown in grey. $\epsilon$ Eridani b is marked as a red circle.}
    \label{fig:ericet}
\end{figure*}

$\tau$ Ceti is another very close, sun-like star and has multiple confirmed planets (\citealt{taucetiplanets1}, \citealt{taucetiplanets2}). However, the known planet with the furthest angular separation is $\tau$ Ceti f, which only reaches $\sim0.37\arcsec$. Currently, these planets would not be detectable, but based on the lower mass limits all being above $1M_\earth$ and Figure \ref{fig:namedfluxes}, these planets will be possible to detect with METIS. Additionally, \cite{dietrich} found that an additional planet could exist within the habitable-zone (likely less massive than the known planets as it has not been detected). Figure \ref{fig:ericet} shows contrast curves for theoretical planets around both $\epsilon$ Eridani and $\tau$ Ceti.

$\epsilon$ Indi is another nearby star with a known exoplanet with a mass of $3.25 M_\jupiter$ and semi-major axis of 11.55 AU \citep{epsindi}. The VLT and LBT currently can reach mass sensitivities of $\sim 10 M_\jupiter$ for $\epsilon$ Indi \citep{siriusnear}. Although currently undetectable, METIS will have the sensitivity to detect it within one hour. 

Procyon presents a similar situation to Sirius with its white dwarf companion at an average separation of roughly 15 AU. We again assume its effects are negligible, but consider the possibility that the thermal evolution of the planets effectively restarted when Procyon B became a white dwarf. Figure \ref{fig:procyon_sens} shows both of these situations with the $\sim$3 SNR curve from NEAR \citep{nearvsmid} in terms of brightness.

$\kappa$ Ceti's habitable zone only extends to 1.5 AU, meaning it would be at a maximum angular separation of 0.17\arcsec, which are not currently possible to resolve. However, this was the youngest FGK type star looked at, allowing for higher internal heat that would make $10M_\jupiter$ planets appear bright ($\sim$1.5 mJy). Such planets are currently detectable with the VLT and LBT.

\subsection{Future Observations}
As seen from Figures \ref{fig:allfluxes} and \ref{fig:namedfluxes}, current 8-meter telescopes, such as the VLT and LBT, typically only have the sensitivity to image Jupiter mass planets in the habitable zone (aside from the closest targets such as $\alpha$ Cen). Therfore, ground-based observations of low mass habitable-zone planets will increase in frequency with the ELTs. Currently, 8-meter telescopes are sensitive to $10M_\jupiter$ habitable-zone planets around close ($\alpha$ Cen, Procyon, etc.) or young ($\kappa$ Ceti, $\beta$ Pic, etc.) stars.

Initial observations with the $JWST$ are exceeding expectations with a brightness sensitivity of $\sim10 \mu$Jy at 1\arcsec \; separation with a SNR of 5 \citep{jwst}. Although only the closest stars have habitable zones at or exceeding that separation, $JWST$ can provide complementary data on low-mass planets on wider orbits.

Other instruments similar to METIS proposed for the next generation of ground-based telescopes are the Thermal Infrared imager for the GMT which provides Extreme contrast and Resolution (TIGER) on the Giant Magellan Telescope (GMT: \citealt{tiger}), and the Mid-IR Camera, High-disperser and IFU spectrograph (MICHI) on the Thirty Meter Telescope (TMT: \citealt{michi}). The $N$-band of for TIGER and MICHI would have nearly the same IWA (0.08\arcsec) and brightness sensitivities. Based on \cite{tiger}, TIGER would have a background-limited sensitivity of $\sim12 \mu$Jy with SNR of 5 for three hour observations for separations greater than 0.2\arcsec. METIS would reach this same sensitivity in two hour observations. These values represent an increase in sensitivity to point sources by a factor of $\sim$100 compared to the VLT and LBT (see also Figure 13 of \citealt{metisiwa}). One hour observations with the next generation of ground-based telescopes also see a factor of $\sim$10 improvement to sensitivity compared to 100 hour VLT sensitivities.
All three instruments could study the entire habitable zones of nearby stars and search for temperate, rocky planets. 

Assuming equal occurrence rates of planets (i.e., based solely on detectablity), we prioritize the brightest targets for observations with METIS or an equivalent instrument mentioned above. If an exo-Earth exists around the closest stars, such as, $\alpha$ Cen A\&B, it would would take three and seven hours to detect at the inner and outer outer edge of the habitable zones for a SNR of 5, respectively. Sirius, $\epsilon$ Eri, and \object{$\tau$ Ceti} would be the other three stars where an exo-Earth could theoretically be imaged at any point in the habitable zone within 100 hours of observation per star with the ELTs.

Instead of observing for 100 hours on a single star, spending only three hours per star will greatly increase stars initially probed, albeit slightly decreasing mass sensitivity. Along with the five stars mentioned above, a habitable-zone Super Earth around Altair or Procyon would be possible to detect in just three hours.

These priority targets are in good agreement with \cite{metisyield}, which instead used HEEPS to generate contrast curves and $Kepler$ occurrence rates to predict the expected yield with METIS. There, the top five candidate were $\alpha$ Cen A\&B, Sirius, Procyon, and Altair. $\tau$ Ceti was included in their 34 hour optimized observation plan (see table 4 therein), but $\epsilon$ Eri was not included as a top potential candidate, despite our finding that a Super Earth in that system could be detected within one hour of observations. However, $\epsilon$ Eri's habitable zone has a relatively small separation from its host star (maximum of 0.33\arcsec) compared to other stars at similar distance. Although the priority targets would not change, the mass sensitivity for three hour observations would shift to Super-Earths ($\sim 10M_\earth$) if METIS projections are overestimated by $\sim$50\%, still representing a factor of 3 increase in mass compared to NEAR results for 100 hour observations.

\section{\label{sec:conc}Conclusions}
In this study, we generated contrast curves for nine different planet masses ranging from one Earth mass to 10 Jupiter masses to predict the theoretical direct detectability in the mid-infrared with current and future ground-based telescopes and to determine optimal initial targets for detecting temperate Earth-sized planets. We looked at 37 FGK type stars within 10 pc and 34 BA type stars within 30 pc. \texttt{AMES-Cond} was used for atmospheric modeling of $5M_\jupiter$ and $10M_\jupiter$ planets while \texttt{petitCODE} was used for the less massive planets. 

While $1M_\jupiter$ habitable-zone planets could only be currently detected around $\alpha$ Cen A and Sirius with typical exposure times of a few nights, we found that $5M_\jupiter$ planets could be imaged in the habitable zone of several nearby stars such as $\alpha$ Cen B, Procyon, Altair, and Vega. For young FGK type stars such as $\epsilon$ Eri and $\tau$ Ceti, improved angular resolution compared to current instruments, such as that provided by mid-infrared imagers on the next generation of  large telescopes (e.g., ELT/METIS) is needed to image habitable-zone planets. Based on detectability alone (i.e., not on occurrence rates), 18 out of 71 stars (10 FGK and 8 BA) could have $10M_\jupiter$ planets that would be bright enough to be detected at 1\arcsec \; separation within 100 hours of observation per star with current instrument sensitivities.

Based on estimated METIS sensitivities for its $N2$-band, 68 of the 71 stars could have their habitable zone imaged with Jupiter-mass planets being detectable within 100 hour observations per system with a SNR of 5 (with many targets requiring far less exposure time). METIS could also begin to image temperate, Earth-sized planets. We found that with 100 hour observations per system, an outer-edge, habitable-zone, Earth-sized planet around Sirius, $\alpha$ Cen A, $\alpha$ Cen B, Procyon, and $\tau$ Ceti could be detected. With only three hours of observation per star, we found that observations $\alpha$ Cen A\&B, Sirius, Procyon, Altair, $\epsilon$ Eri, and $\tau$ Ceti would all be sensitive to a super-Earth in the habitable zone. 

\section*{Acknowledgements}
The results reported herein benefited from collaborations and/or information exchange within the program “Alien Earths” (supported by the National Aeronautics and Space Administration under Agreement No. 80NSSC21K0593) for NASA’s Nexus for Exoplanet System Science (NExSS) research coordination network sponsored by NASA’s Science Mission Directorate. K.W. acknowledges support from NASA through the NASA Hubble Fellowship grant HST-HF2-51472.001-A awarded by the Space Telescope Science Institute, which is operated by the Association of Universities for Research in Astronomy, Incorporated, under NASA contract NAS5-26555. We acknowledge the use of the software packages SciPy \citep{Virtanen_2020}, NumPy \citep{harris2020array}, matplotlib \citep{Hunter:2007}, and pandas \citep{McKinney_2010, McKinney_2011}. The citations in this paper have made use of NASA’s Astrophysics Data System Bibliographic Services.

\bibliography{midbir}

\appendix
\section{Complete Data on all Analyzed Stars}

\begin{deluxetable*}{cccccc}
\tablenum{1}
\tablecaption{General information on all FGK type stars analyzed\label{tab:fgktable}}
\tablewidth{0pt}
\tablehead{
\colhead{Henry} & \colhead{Common Name} & \colhead{Spectral} & \colhead{Distance} &
\colhead{Minimum} & \colhead{Maximum} \\
\colhead{Draper} & \colhead{Or Designation} & \colhead{Type} & \colhead{(pc)} &
\colhead{Age (Myr)} & \colhead{Age (Myr)}
}
\decimalcolnumbers
\startdata
128620       & $\alpha$ Cen A   & G2V           & 1.35    & 5$^a$                     & 5.6$^a$              \\
128621       & $\alpha$ Cen B   & K1V           & 1.35    & 5$^a$     & 5.6$^a$            \\
22049        & $\epsilon$ Eri   & K2V           & 3.22    & 0.4$^b$                                                        & 0.8$^b$        \\
61421        & Procyon          & F5V           & 3.50    & 1.74$^c$                                                       & 2$^c$             \\
209100       & $\epsilon$ Ind   & K5V           & 3.63    & 3.7$^d$                                                        & 5.7$^d$            \\
10700        & $\tau$ Cet       & G8V           & 3.65    &                      & 5.8$^b$                      \\
88230        & Groombridge 1618 & K8V           & 4.87    &                       & 6.6$^e$                         \\
26965        & $o$ Eri   & K1V           & 5.04    & 4.3$^b$                        & 5.6$^b$                  \\
165341       & 70 Oph           & K0V           & 5.09     & 1.1$^b$                                                        & 1.9$^b$             \\
185144       & $\sigma$ Dra     & K0V           & 5.77    & 2.4$^f$                                                       & 3.6$^f$           \\
4614         & $\eta$ Cas       & G0V           & 5.95    & 4.5$^g$                                                        & 6.3$^g$        \\
191408       & Gliese 783 A     & K2V           & 6.05    & 6.4$^b$                                                        & 7.7$^b$             \\
20794        & 82 G. Eri        & G8V           & 6.06    & 6.1$^b$                                                        & 6.6$^b$               \\
79211        & Gliese 338 B     & K2            & 6.27    & 1$^h$                & 7$^h$                     \\
131156       & $\xi$ Boo        & G8V           & 6.70    &                      & 2$^i$                            \\
4628         & 96 G. Psc        & K2V           & 7.46    & 4$^b$                       & 5.4$^b$          \\
10476        & 107 Psc          & K1V           & 7.47    & 5$^b$                      & 6.3$^b$          \\
6582         & $\mu$ Cas        & G5Vp          & 7.55    & 5.3$^b$                                                        & 5.9$^b$   \\
30652        & $\pi^3$ Ori      & F6V           & 8.03    &                 & 1.4$^j$                       \\
170153       & $\chi$ Dra       & F7Vvar        & 8.06    &            & 5.3$^j$                      \\
10360        & p Eridani        & K0V           & 8.15    & 4.8$^b$                                                        & 6.2$^b$     \\
109358       & $\beta$ CVn      & G0V           & 8.37    & 3.3$^b$                                                        & 6.4$^b$   \\
115617       & 61 Vir           & G5V           & 8.53    & 6.1$^b$                                                        & 6.6$^b$    \\
1581         & $\zeta$ Tuc      & F9V           & 8.59    & 2.1$^b$                                                        & 3.02$^k$   \\
39587        & $\chi$ Ori       & G0V           & 8.66    & 0.3$^b$                                                        & 0.4$^b$    \\
50281        & Gliese 250 A     & K3V           & 8.70     & 1.22$^l$                                                       & 4.22$^l$   \\
32147        & Gliese 183       & K3V           & 8.81    & 2$^m$            & 4.5$^n$           \\
192310       & Gliese 785       & K3V           & 8.82    & 7.5$^b$                                                        & 8.9$^b$   \\
38393        & $\gamma$ Lep     & F7V           & 8.97     & 1.3$^j$                                                        & 1.3$^j$   \\
114710       & $\beta$ Com      & G0V           & 9.15    & 1.5$^b$                                                        & 2.5$^b$  \\
103095       & Groombridge 1830 & G8Vp          & 9.16    & 4.7$^b$                                                        & 5.3$^b$\\
20630        & $\kappa$ Cet     & G5Vvar        & 9.16    & 0.3$^b$                                                        & 0.4$^b$ \\
203608       & $\gamma$ Pav     & F6V           & 9.22   & 1$^j$                      & 7.25$^o$\\
102365       & G. Cen           & G2V           & 9.24   & 4.5$^b$                                                        & 5.7$^b$\\
101501       & 61 UMa           & G8V           & 9.54    & 0.4$^p$                                                        & 3.8$^p$  \\
149661       & 12 Oph           & K2V           & 9.78    & 1$^b$            & 1.9$^b$ \\
10780        & Gliese 75        & K0V           & 9.98    & 1.8$^b$                                                        & 2.9$^b$ \\ 
\enddata
\tablecomments{Missing value indicates no minimum age was listed. Age sources: a: \cite{alphaceninfo}. b: \cite{agesource}. c: \cite{procyonage}. d: \cite{indimass}. e: \cite{groomage}. f: \cite{sigmadraage}. g: \cite{etacasage}. h: \cite{gliese338age}. i: \cite{xibooage}. j: \cite{pioriage}. k: \cite{maxzetatuc}. l: \cite{gliese250age}. m: \cite{gliese183min}. n: \cite{gliese183age}. o: \cite{gampavmax}. p: \cite{umaage}.}
\end{deluxetable*}

\begin{deluxetable*}{cccccch}
\tablenum{2}
\tablecaption{General information on all BA type stars analyzed\label{tab:batable}}
\tablewidth{0pt}
\tablehead{
\colhead{Henry} & \colhead{Common Name} & \colhead{Spectral} & \colhead{Distance} &
\colhead{Minimum} & \colhead{Maximum} & \nocolhead{V} \\
\colhead{Draper} & \colhead{Or Designation} & \colhead{Type} & \colhead{(pc)} &
\colhead{Age (Myr)} & \colhead{Age (Myr)} & \nocolhead{(mag)}
}
\decimalcolnumbers
\startdata
48915        & Sirius     & A0V           & 2.64    & 237$^a$               & 247$^a$               & 88.2   \\
187642       & Altair     & A7V           & 5.14     & 285$^b$               & 1055$^b$              & 9.89   \\
216956       & Fomalhaut  & A3V           & 7.69    & 455$^b$               & 616$^b$               & 10.1    \\
172167       & Vega       & A0V           & 7.76    & 366$^b$               & 500$^b$               & 11.0    \\
102647       & Denebola   & A3V           & 11.1    & 247$^b$               & 593$^b$               & 10.9  \\
203280       & Alderamin  & A7V           & 15.0    & 813$^b$               & 1132$^b$              & 1.52    \\
97603        & $\delta$ Leo          & A4V           & 17.7    & 602$^b$               & 865$^b$               & 2.68         \\
115892       & $\iota$ Cen          & A2V           & 18.0    & 107$^b$               & 413$^b$               & 13.1         \\
11636        & $\beta$ Ari          & A5V        & 18.3    & 163$^c$               & 649$^c$               & 14.1         \\
39060        & $\beta$ Pic          & A3V           & 19.3    & 8$^d$                 & 21$^d$                & 228          \\
38678        & $\zeta$ Lep          & A2V        & 21.5    & 376$^b$               & 750$^b$               & 7.35          \\
118098       & $\zeta$ Vir          & A3V           & 22.4    & 521$^b$               & 827$^b$               & 3.00         \\
139006       & Alphekka   & A0V           & 22.9    & 263$^b$               & 399$^b$               & 5.03      \\
2262         & $\kappa$ Phe          & A7V           & 23.5    & 266$^b$               & 866$^b$               & 4.00       \\
87901        & Regulus    & B7V           & 23.8    & 92$^b$                & 129$^b$               & 7.85          \\
116656       & Mizar      & A2V           & 24.0    & 97$^b$                & 398$^b$               & 10.1       \\
95418        & Merak      & A1V           & 24.3     & 173$^b$               & 501$^b$               & 4.32       \\
112185       & Alioth     & A0p           & 24.8    & 266$^b$               & 501$^b$               & 2.29     \\
106591       & $\delta$ UMa          & A3Vvar        & 25.0     & 87$^b$                & 377$^b$               & 30.9            \\
16970        & $\gamma$ Cet          & A3V           & 25.1    & 463$^e$               & 751$^e$               & 3.93         \\
40183        & $\beta$ Aur          & A2V           & 25.2    & 441$^b$               & 569$^b$               & 3.24          \\
177724       & $\zeta$ Aql          & A0Vn          & 25.5    & 211$^b$               & 259$^b$               & 11.1       \\
103287       & Phad       & A0V         & 25.6     & 164$^b$               & 421$^b$               & 4.97     \\
99211        & $\gamma$ Crt          & A9V           & 25.7    & 635$^b$               & 907$^b$               & 5.72       \\
18978        & $\tau$ Eri          & A4V           & 26.4    & 428$^b$               & 926$^b$               & 3.54        \\
108767       & $\delta$ Crv          & B9.5V         & 26.9    & 168$^b$               & 289$^b$               & 11.7            \\
87696        & 21 LMi          & A7V           & 27.9    & 596$^b$               & 972$^b$               & 3.11         \\
19356        & Algol      & B8V           & 28.5    & 125$^b$               & 163$^b$               & 15.0     \\
70060        & GJ 1109          & A4m        & 28.5    & 349$^b$               & 999$^b$               & 4.04     \\
56537        & $\lambda$ Gem          & A3V        & 28.9    & 116$^b$               & 488$^b$               & 5.62           \\
161868       & $\gamma$ Oph          & A0V           & 29.1    & 182$^b$               & 426$^b$               & 12.9      \\
135379       & $\beta$ Cir          & A3V           & 29.6    & 449$^b$               & 677$^b$               & 8.84         \\
358          & Alpheratz  & B9p           & 29.8    & 60$^f$                & 70$^f$                & 29.0           \\
125161       & $\iota$ Boo          & A9V           & 29.8    & 345$^b$               & 1000$^b$              & 5.18            \\
\enddata
\tablecomments{Age sources: a: \cite{siriusage}. b: \cite{altairage}. c: \cite{betaariage}. d: \cite{betapicage}. e: \cite{gammacetage}. f: \cite{alpheratzage}.}
\end{deluxetable*}

\begin{deluxetable*}{ccccccc}
\tablenum{3}
\tablecaption{Fluxes and Contrasts of various Jupiter masses around FGK type stars \label{tab:fgkfluxes}}
\tablewidth{0pt}
\tablehead{
\colhead{Name} & \colhead{1 M$_\jupiter$ Contrast} & \colhead{1 M$_\jupiter$ Flux} & \colhead{5 M$_\jupiter$ Contrast} &
\colhead{5 M$_\jupiter$ Flux} & \colhead{10 M$_\jupiter$ Contrast} & \colhead{10 M$_\jupiter$ Flux} \\
\colhead{} & \colhead{at 1\arcsec} & \colhead{at 1\arcsec (mJy)} & \colhead{at 1\arcsec} &
\colhead{at 1\arcsec (mJy)} & \colhead{at 1\arcsec} & \colhead{at 1\arcsec (mJy)}
}
\decimalcolnumbers
\startdata
$\alpha$ Cen A & 8.827e-06& 1.960e+00 &1.875e-05 &4.162e+00& 4.595e-05 &1.020e+01  \\
$\alpha$ Cen B & 4.704e-06& 2.841e-01 &2.224e-05 &1.343e+00& 8.678e-05 &5.242e+00  \\
$\epsilon$ Eri & 1.150e-05& 1.095e-01 &2.768e-04 &2.635e+00& 8.205e-04 &7.811e+00  \\
Procyon & 1.987e-06& 1.572e-01 &9.623e-06 &7.612e-01& 2.855e-05 &2.258e+00  \\
$\epsilon$ Ind & 3.946e-07& 2.210e-03 &2.077e-05 &1.163e-01& 1.253e-04 &7.017e-01  \\
 $\tau$ Cet & 3.912e-07& 3.740e-03 &1.031e-05 &9.856e-02& 6.619e-05 &6.328e-01  \\
 Groombridge 1618 & 0.000e+00& 0.000e+00 &1.512e-05 &2.601e-02& 1.231e-04 &2.117e-01  \\
$o$ Eri  & 2.844e-07& 1.413e-03 &1.411e-05 &7.013e-02& 8.471e-05 &4.210e-01  \\
70 Oph & 1.308e-06& 9.640e-03 &1.049e-04 &7.731e-01& 3.307e-04 &2.437e+00  \\
 $\sigma$ Dra & 5.786e-07& 1.915e-03 &3.625e-05 &1.200e-01& 1.632e-04 &5.402e-01  \\
 $\eta$ Cas & 2.087e-07& 1.592e-03 &7.703e-06 &5.877e-02& 4.465e-05 &3.407e-01  \\
Gliese 783 & 0.000e+00& 0.000e+00 &1.032e-05 &2.694e-02& 8.276e-05 &2.160e-01  \\
82 G. Eri  & 1.467e-07& 6.411e-04 &7.048e-06 &3.080e-02& 4.749e-05 &2.075e-01  \\
Gliese 338 B & 4.999e-07& 4.254e-04 &5.367e-05 &4.567e-02& 2.884e-04 &2.454e-01  \\
$\xi$ Boo & 1.003e-06& 3.912e-03 &5.289e-05 &2.063e-01& 2.053e-04 &8.007e-01  \\
 96 G. Psc & 2.315e-07& 3.611e-04 &1.666e-05 &2.599e-02& 1.028e-04 &1.604e-01  \\
107 Psc & 1.703e-07& 3.406e-04 &9.294e-06 &1.859e-02& 6.447e-05 &1.289e-01  \\
$\mu$ Cas & 1.699e-07& 3.432e-04 &9.260e-06 &1.871e-02& 6.422e-05 &1.297e-01  \\
$\pi^3$ Ori & 7.000e-07& 4.522e-03 &3.914e-05 &2.528e-01& 1.227e-04 &7.926e-01  \\
$\chi$ Dra & 1.250e-07& 7.087e-04 &5.339e-06 &3.027e-02& 3.137e-05 &1.779e-01  \\
p Eridani & 1.836e-07& 4.920e-04 &1.584e-05 &4.245e-02& 9.786e-05 &2.623e-01  \\
$\beta$ CVn & 1.214e-07& 4.115e-04 &6.338e-06 &2.149e-02& 3.821e-05 &1.295e-01  \\
61 Vir    & 1.015e-07& 2.598e-04 &6.015e-06 &1.540e-02& 4.128e-05 &1.057e-01  \\
$\zeta$ Tuc   & 3.412e-07& 1.088e-03 &1.627e-05 &5.190e-02& 7.281e-05 &2.323e-01  \\
$\chi$ Ori   & 1.325e-05& 3.816e-02 &2.549e-04 &7.341e-01& 6.555e-04 &1.888e+00  \\
Gliese 250 A  & 7.424e-07& 6.786e-04 &4.683e-05 &4.280e-02& 2.128e-04 &1.945e-01  \\
Gliese 183   & 4.327e-07& 5.668e-04 &4.108e-05 &5.381e-02& 1.864e-04 &2.442e-01  \\
Gliese 785   & 0.000e+00& 0.000e+00 &4.754e-06 &7.797e-03& 4.581e-05 &7.513e-02  \\
$\gamma$ Lep & 7.069e-07& 3.110e-03 &3.841e-05 &1.690e-01& 1.208e-04 &5.315e-01  \\
$\beta$ Com & 5.230e-07& 1.511e-03 &2.631e-05 &7.604e-02& 1.019e-04 &2.945e-01  \\
Groombridge 1830 & 2.266e-07& 2.080e-04 &1.834e-05 &1.684e-02& 1.140e-04 &1.047e-01  \\
$\kappa$ Cet  & 1.477e-05& 3.013e-02 &2.862e-04 &5.838e-01& 7.361e-04 &1.502e+00  \\
$\gamma$ Pav & 1.342e-07& 3.972e-04 &8.576e-06 &2.538e-02& 4.484e-05 &1.327e-01  \\
G. Cen & 1.415e-07& 2.788e-04 &8.968e-06 &1.767e-02& 5.485e-05 &1.081e-01  \\
61 UMa & 7.829e-07& 2.012e-03 &4.752e-05 &1.221e-01& 1.851e-04 &4.757e-01  \\
12 Oph & 1.211e-06& 1.780e-03 &1.198e-04 &1.761e-01& 3.792e-04 &5.574e-01  \\
Gliese 75  & 7.623e-07& 6.983e-04 &5.455e-05 &4.997e-02& 2.127e-04 &1.948e-01  \\
\enddata

\end{deluxetable*}

\begin{deluxetable*}{ccccccc}
\tablenum{4}
\tablecaption{Fluxes and Contrasts of various Jupiter masses around BA type stars \label{tab:bafluxes}}
\tablewidth{0pt}
\tablehead{
\colhead{Name} & \colhead{1 M$_\jupiter$ Contrast} & \colhead{1 M$_\jupiter$ Flux} & \colhead{5 M$_\jupiter$ Contrast} &
\colhead{5 M$_\jupiter$ Flux} & \colhead{10 M$_\jupiter$ Contrast} & \colhead{10 M$_\jupiter$ Flux} \\
\colhead{} & \colhead{at 1\arcsec} & \colhead{at 1\arcsec (mJy)} & \colhead{at 1\arcsec} &
\colhead{at 1\arcsec (mJy)} & \colhead{at 1\arcsec} & \colhead{at 1\arcsec (mJy)}
}
\decimalcolnumbers
\startdata
Sirius & 2.119e-05& 3.306e+00 &8.522e-05 &1.329e+01& 1.861e-04 &2.903e+01  \\
Altair & 2.388e-06& 7.880e-02 &2.148e-05 &7.088e-01& 5.995e-05 &1.978e+00  \\
Fomalhaut & 2.461e-06& 4.430e-02 &3.192e-05 &5.746e-01& 8.732e-05 &1.572e+00  \\
Vega  & 2.640e-06& 1.098e-01 &1.866e-05 &7.763e-01& 4.887e-05 &2.033e+00  \\
Denebola & 2.617e-06& 1.824e-02 &4.136e-05 &2.883e-01& 1.145e-04 &7.981e-01  \\
Alderamin & 3.672e-07& 2.570e-03 &7.429e-06 &5.200e-02& 2.291e-05 &1.604e-01  \\
$\delta$ Leo & 6.453e-07& 3.530e-03 &1.605e-05 &8.779e-02& 4.750e-05 &2.598e-01  \\
$\iota$ Cen & 3.140e-06& 1.118e-02 &3.924e-05 &1.397e-01& 1.007e-04 &3.585e-01  \\
$\beta$ Ari  & 3.404e-06& 1.603e-02 &5.597e-05 &2.636e-01& 1.558e-04 &7.338e-01  \\
$\beta$ Pic   & 5.491e-05& 1.861e-01 &6.338e-04 &2.149e+00& 1.354e-03 &4.590e+00  \\
$\zeta$ Lep  & 1.776e-06& 3.765e-03 &3.726e-05 &7.899e-02& 1.103e-04 &2.338e-01  \\
$\zeta$ Vir  & 7.227e-07& 1.987e-03 &1.690e-05 &4.648e-02& 5.024e-05 &1.382e-01  \\
Alphekka & 1.213e-06& 7.048e-03 &1.818e-05 &1.056e-01& 4.646e-05 &2.699e-01  \\
$\kappa$ Phe & 9.631e-07& 1.502e-03 &2.154e-05 &3.360e-02& 6.395e-05 &9.976e-02  \\
Regulus  & 1.895e-06& 1.738e-02 &1.746e-05 &1.601e-01& 4.052e-05 &3.716e-01  \\
Mizar& 2.415e-06& 1.584e-02 &3.929e-05 &2.577e-01& 9.505e-05 &6.235e-01  \\
Merak & 1.038e-06& 5.024e-03 &1.732e-05 &8.383e-02& 4.433e-05 &2.146e-01  \\
Alioth & 5.485e-07& 4.766e-03 &6.776e-06 &5.888e-02& 1.868e-05 &1.623e-01  \\
$\delta$ UMa & 7.445e-06& 1.794e-02 &1.174e-04 &2.829e-01& 2.844e-04 &6.854e-01  \\
$\gamma$ Cet & 9.502e-07& 2.014e-03 &2.317e-05 &4.912e-02& 6.874e-05 &1.457e-01  \\
$\beta$ Aur  & 7.812e-07& 6.125e-03 &1.310e-05 &1.027e-01& 3.657e-05 &2.867e-01  \\
$\zeta$ Aql & 2.679e-06& 7.716e-03 &4.344e-05 &1.251e-01& 1.051e-04 &3.027e-01  \\
Phad    & 1.194e-06& 5.385e-03 &1.728e-05 &7.793e-02& 4.426e-05 &1.996e-01  \\
$\gamma$ Crt & 1.376e-06& 2.133e-03 &4.315e-05 &6.688e-02& 1.285e-04 &1.992e-01  \\
$\tau$ Eri & 8.557e-07& 1.164e-03 &2.110e-05 &2.870e-02& 6.288e-05 &8.552e-02  \\
$\delta$ Crv & 2.817e-06& 7.212e-03 &4.193e-05 &1.073e-01& 1.014e-04 &2.596e-01  \\
21 LMi   & 7.500e-07& 7.875e-04 &2.308e-05 &2.423e-02& 6.871e-05 &7.215e-02  \\
Algol  & 3.606e-06& 2.643e-02 &3.734e-05 &2.737e-01& 8.903e-05 &6.526e-01  \\
GJ 1109  & 9.753e-07& 1.112e-03 &2.472e-05 &2.818e-02& 7.372e-05 &8.404e-02  \\
$\lambda$ Gem & 1.359e-06& 2.704e-03 &2.141e-05 &4.261e-02& 5.501e-05 &1.095e-01  \\
$\gamma$ Oph & 3.096e-06& 4.458e-03 &4.908e-05 &7.068e-02& 1.262e-04 &1.817e-01  \\
$\beta$ Cir& 2.127e-06& 2.893e-03 &4.904e-05 &6.669e-02& 1.458e-04 &1.983e-01  \\
Alpheratz& 6.992e-06& 3.342e-02 &6.919e-05 &3.307e-01& 1.607e-04 &7.681e-01  \\
$\iota$ Boo & 1.247e-06& 1.079e-03 &3.210e-05 &2.777e-02& 9.580e-05 &8.287e-02  \\
\enddata

\end{deluxetable*}

\begin{deluxetable*}{ccccccc}
\tablenum{5}
\tablecaption{Fluxes and Contrasts of various Earth masses around FGK type stars \label{tab:fgkearthfluxes}}
\tablewidth{0pt}
\tablehead{
\colhead{Name} & \colhead{1 M$_\earth$ Contrast} & \colhead{1 M$_\earth$ Flux} & \colhead{5 M$_\earth$ Contrast} &
\colhead{5 M$_\earth$ Flux} & \colhead{10 M$_\earth$ Contrast} & \colhead{10 M$_\earth$ Flux} \\
\colhead{} & \colhead{at 1\arcsec} & \colhead{at 1\arcsec (mJy)} & \colhead{at 1\arcsec} &
\colhead{at 1\arcsec (mJy)} & \colhead{at 1\arcsec} & \colhead{at 1\arcsec (mJy)}
}
\decimalcolnumbers
\startdata
$\alpha$ Cen A & 1.287e-07& 2.857e-02& 5.036e-07 &1.118e-01 & 1.134e-06 & 2.517e-01 \\
$\alpha$ Cen B & 6.300e-08& 3.805e-03& 2.140e-07 &1.293e-02 & 4.829e-07 & 2.917e-02 \\
Procyon & 2.286e-08& 1.808e-03& 1.142e-07 &9.033e-03 & 2.288e-07 & 1.810e-02 \\
\enddata

\end{deluxetable*}

\begin{deluxetable*}{ccccccc}
\tablenum{6}
\tablecaption{Fluxes and Contrasts of various Earth masses around BA type stars \label{tab:baearthfluxes}}
\tablewidth{0pt}
\tablehead{
\colhead{Name} & \colhead{1 M$_\earth$ Contrast} & \colhead{1 M$_\earth$ Flux} & \colhead{5 M$_\earth$ Contrast} &
\colhead{5 M$_\earth$ Flux} & \colhead{10 M$_\earth$ Contrast} & \colhead{10 M$_\earth$ Flux} \\
\colhead{} & \colhead{at 1\arcsec} & \colhead{at 1\arcsec (mJy)} & \colhead{at 1\arcsec} &
\colhead{at 1\arcsec (mJy)} & \colhead{at 1\arcsec} & \colhead{at 1\arcsec (mJy)}
}
\decimalcolnumbers
\startdata
Sirius & 1.842e-07& 2.874e-02& 1.737e-06 &2.710e-01 & 3.269e-06 & 5.100e-01 \\
Altair & 1.498e-08& 4.943e-04& 8.841e-08 &2.918e-03 & 1.755e-07 & 5.792e-03 \\
Fomalhaut & 8.088e-09& 1.456e-04& 4.730e-08 &8.514e-04 & 9.405e-08 & 1.693e-03 \\
Vega  & 1.736e-08& 7.222e-04& 1.182e-07 &4.917e-03 & 2.376e-07 & 9.884e-03 \\
Denebola & 1.731e-09& 1.207e-05& 1.009e-08 &7.033e-05 & 2.115e-08 & 1.474e-04 \\
Alderamin & 7.379e-10& 5.165e-06& 3.263e-09 &2.284e-05 & 6.983e-09 & 4.888e-05 \\
$\delta$ Leo & 1.326e-09& 7.253e-06& 6.452e-09 &3.529e-05 & 1.314e-08 & 7.188e-05 \\
$\iota$ Cen & 3.193e-10& 1.747e-06& 1.553e-09 &8.495e-06 & 3.162e-09 & 1.730e-05 \\
$\beta$ Pic   & 0.000e+00& 0.000e+00& 3.678e-08 &1.247e-04 & 1.571e-07 & 5.326e-04 \\
Alphekka & 7.772e-10& 4.516e-06& 5.014e-09 &2.913e-05 & 1.036e-08 & 6.019e-05 \\
Regulus  & 3.206e-09& 2.940e-05& 3.384e-08 &3.103e-04 & 6.546e-08 & 6.003e-04 \\
Mizar & 2.161e-10& 1.418e-06& 1.815e-09 &1.191e-05 & 5.199e-09 & 3.411e-05 \\
Merak & 4.782e-10& 2.314e-06& 3.094e-09 &1.497e-05 & 6.344e-09 & 3.070e-05 \\
Alioth & 5.752e-10& 4.998e-06& 3.478e-09 &3.022e-05 & 7.074e-09 & 6.147e-05 \\
$\beta$ Aur  & 3.571e-10& 2.800e-06& 1.967e-09 &1.542e-05 & 3.930e-09 & 3.081e-05 \\
$\zeta$ Aql & 2.552e-10& 7.350e-07& 2.225e-09 &6.408e-06 & 6.068e-09 & 1.748e-05 \\
Phad    & 3.896e-10& 1.757e-06& 2.735e-09 &1.233e-05 & 5.921e-09 & 2.670e-05 \\
$\delta$ Crv & 4.120e-10& 1.055e-06& 3.401e-09 &8.707e-06 & 8.542e-09 & 2.187e-05 \\
Algol  & 2.517e-09& 1.845e-05& 2.228e-08 &1.633e-04 & 4.977e-08 & 3.648e-04 \\
Alpheratz  & 2.949e-09& 1.410e-05& 3.997e-08 &1.911e-04 & 8.112e-08 & 3.878e-04 \\
\enddata

\end{deluxetable*}

\begin{deluxetable*}{ccccccccc}
\tablenum{7}
\tablecaption{Fluxes of $1 M_\earth$, $10 M_\earth$, and $1 M_\jupiter$ planets at the inner and outer edge of the star's habitable zones for FGK type stars. \label{tab:fgkhabzone}}
\tablewidth{0pt}
\tablehead{
\colhead{Name} & \colhead{inner HZ} & \colhead{1 M$_\earth$ Flux} & \colhead{10 M$_\earth$ Flux} & \colhead{1 M$_\jupiter$ Flux} &
\colhead{Outer HZ} & \colhead{1 M$_\earth$ Flux}&  \colhead{10 M$_\earth$ Flux} & \colhead{1 M$_\jupiter$ Flux} \\
\colhead{} & \colhead{(arcsec)} & \colhead{(mJy)} & \colhead{(mJy)} & \colhead{(mJy)} &
\colhead{(arcsec)} & \colhead{(mJy)} & \colhead{(mJy)} & \colhead{(mJy)}
}
\decimalcolnumbers
\startdata
$\alpha$ Cen A & 0.86& 4.311e-02 &3.965e-01 &2.993e+00& 1.52 &9.320e-03 &7.839e-02 &6.662e-01 \\
$\alpha$ Cen B & 0.511& 2.813e-02 &2.624e-01 &1.970e+00& 0.92 &5.355e-03 &4.828e-02 &3.854e-01 \\
$\epsilon$ Eri & 0.182& 3.759e-03 &5.007e-02 &4.571e-01& 0.326 &1.888e-03 &2.794e-02 &2.978e-01 \\
Procyon & 0.684& 5.053e-03 &5.586e-02 &4.042e-01& 1.187 &1.164e-03 &1.209e-02 &1.098e-01 \\
$\epsilon$ Ind & 0.127& 1.385e-03 &1.218e-02 &9.867e-02& 0.234 &9.789e-04 &6.647e-03 &7.123e-02 \\
 $\tau$ Cet & 0.193& 3.187e-03 &2.945e-02 &2.245e-01& 0.344 &1.347e-03 &1.645e-02 &9.544e-02 \\
 Groombridge 1618 & 0.082& 7.735e-05 &6.766e-04 &6.500e-03& 0.155 &7.281e-05 &6.766e-04 &4.971e-03 \\
$o$ Eri  & 0.131& 9.562e-04 &8.504e-03 &6.779e-02& 0.234 &6.883e-04 &4.710e-03 &4.969e-02 \\
70 Oph & 0.147& 1.417e-03 &1.549e-02 &1.262e-01& 0.264 &1.000e-03 &8.593e-03 &9.367e-02 \\
 $\sigma$ Dra & 0.109& 4.488e-04 &4.081e-03 &3.667e-02& 0.195 &3.290e-04 &2.393e-03 &2.798e-02 \\
 $\eta$ Cas & 0.175& 1.425e-03 &1.319e-02 &1.003e-01& 0.307 &9.148e-04 &7.490e-03 &6.472e-02 \\
Gliese 783 A  & 0.083& 2.545e-04 &1.928e-03 &1.778e-02& 0.15 &1.915e-04 &1.432e-03 &1.370e-02 \\
82 G. Eri  & 0.138& 5.572e-04 &4.785e-03 &3.846e-02& 0.246 &3.988e-04 &2.654e-03 &2.795e-02 \\
Gliese 338 B & 0.046& 2.177e-05 &2.076e-04 &2.584e-03& 0.087 &2.177e-05 &2.076e-04 &2.217e-03 \\
$\xi$ Boo & 0.108& 4.696e-04 &4.727e-03 &4.321e-02& 0.191 &3.476e-04 &2.710e-03 &3.382e-02 \\
 96 G. Psc & 0.07& 9.667e-05 &7.632e-04 &7.644e-03& 0.127 &7.488e-05 &5.976e-04 &6.142e-03 \\
107 Psc & 0.089& 1.371e-04 &1.099e-03 &1.016e-02& 0.161 &1.028e-04 &7.920e-04 &7.856e-03 \\
$\mu$ Cas & 0.086& 1.406e-04 &1.129e-03 &1.040e-02& 0.155 &1.059e-04 &7.975e-04 &8.074e-03 \\
$\pi^3$ Ori & 0.19& 8.463e-04 &9.735e-03 &7.268e-02& 0.331 &4.377e-04 &5.587e-03 &4.028e-02 \\
$\chi$ Dra & 0.157& 6.237e-04 &5.681e-03 &4.390e-02& 0.275 &4.471e-04 &3.226e-03 &3.177e-02 \\
p Eridani & 0.071& 1.291e-04 &9.884e-04 &1.027e-02& 0.128 &9.913e-05 &8.399e-04 &8.195e-03 \\
$\beta$ CVn & 0.123& 2.636e-04 &2.320e-03 &1.874e-02& 0.217 &1.925e-04 &1.291e-03 &1.395e-02 \\
61 Vir    & 0.103& 1.821e-04 &1.498e-03 &1.271e-02& 0.182 &1.353e-04 &9.101e-04 &9.631e-03 \\
$\zeta$ Tuc   & 0.123& 2.962e-04 &2.956e-03 &2.392e-02& 0.215 &2.183e-04 &1.661e-03 &1.821e-02 \\
$\chi$ Ori   & 0.113& 2.574e-04 &3.614e-03 &6.800e-02& 0.198 &1.907e-04 &2.039e-03 &6.137e-02 \\
Gliese 250 A  & 0.055& 2.428e-05 &2.182e-04 &3.456e-03& 0.1 &2.105e-05 &2.182e-04 &2.970e-03 \\
Gliese 183   & 0.063& 4.902e-05 &3.947e-04 &5.096e-03& 0.114 &3.827e-05 &3.713e-04 &4.234e-03 \\
Gliese 785   & 0.07& 6.652e-05 &4.689e-04 &4.646e-03& 0.125 &5.146e-05 &4.049e-04 &3.685e-03 \\
$\gamma$ Lep & 0.155& 4.311e-04 &4.906e-03 &3.855e-02& 0.271 &3.116e-04 &2.802e-03 &2.917e-02 \\
$\beta$ Com & 0.12& 2.263e-04 &2.330e-03 &1.992e-02& 0.211 &1.666e-04 &1.306e-03 &1.539e-02 \\
Groombridge 1830 & 0.05& 1.669e-05 &1.531e-04 &1.820e-03& 0.092 &1.637e-05 &1.531e-04 &1.545e-03 \\
$\kappa$ Cet  & 0.096& 1.382e-04 &1.898e-03 &4.800e-02& 0.17 &1.041e-04 &1.060e-03 &4.435e-02 \\
$\gamma$ Pav & 0.125& 2.531e-04 &2.335e-03 &1.828e-02& 0.218 &1.864e-04 &1.311e-03 &1.369e-02 \\
G. Cen & 0.096& 1.235e-04 &1.017e-03 &9.058e-03& 0.169 &9.275e-05 &6.493e-04 &6.992e-03 \\
61 UMa & 0.079& 1.254e-04 &1.148e-03 &1.388e-02& 0.141 &9.594e-05 &8.566e-04 &1.144e-02 \\
12 Oph & 0.062& 5.968e-05 &5.674e-04 &8.273e-03& 0.112 &4.713e-05 &4.463e-04 &7.130e-03 \\
Gliese 75  & 0.07& 3.639e-05 &3.235e-04 &4.338e-03& 0.125 &2.822e-05 &2.725e-04 &3.637e-03 \\
\enddata

\end{deluxetable*}

\begin{deluxetable*}{ccccccccc}
\tablenum{8}
\tablecaption{Fluxes of $1 M_\earth$, $10 M_\earth$, and $1 M_\jupiter$ planets at the inner and outer edge of the star's habitable zones for BA type stars. \label{tab:bahabzone}}
\tablewidth{0pt}
\tablehead{
\colhead{Name} & \colhead{inner HZ} & \colhead{1 M$_\earth$ Flux} & \colhead{10 M$_\earth$ Flux} & \colhead{1 M$_\jupiter$ Flux} &
\colhead{Outer HZ} & \colhead{1 M$_\earth$ Flux}&  \colhead{10 M$_\earth$ Flux} & \colhead{1 M$_\jupiter$ Flux} \\
\colhead{} & \colhead{(arcsec)} & \colhead{(mJy)} & \colhead{(mJy)} & \colhead{(mJy)} &
\colhead{(arcsec)} & \colhead{(mJy)} & \colhead{(mJy)} & \colhead{(mJy)}
}
\decimalcolnumbers
\startdata
Sirius & 2.16& 5.778e-03 &9.190e-02 &1.302e+00& 3.0 &2.599e-03 &4.235e-02 &1.023e+00 \\
Altair & 0.549& 2.483e-03 &3.356e-02 &2.414e-01& 0.946 &6.184e-04 &8.204e-03 &8.884e-02 \\
Fomalhaut & 0.465& 1.378e-03 &1.928e-02 &1.475e-01& 0.802 &3.141e-04 &4.496e-03 &5.900e-02 \\
Vega  & 0.829& 1.191e-03 &1.707e-02 &1.491e-01& 1.446 &2.775e-04 &3.714e-03 &7.188e-02 \\
Denebola & 0.305& 7.235e-04 &1.108e-02 &8.176e-02& 0.525 &1.391e-04 &2.674e-03 &3.084e-02 \\
Alderamin & 0.238& 3.788e-04 &4.965e-03 &3.387e-02& 0.41 &1.290e-04 &2.669e-03 &1.285e-02 \\
$\delta$ Leo & 0.193& 2.885e-04 &3.838e-03 &2.811e-02& 0.332 &1.553e-04 &2.245e-03 &1.699e-02 \\
$\iota$ Cen & 0.249& 1.745e-04 &2.897e-03 &2.758e-02& 0.429 &4.432e-05 &1.243e-03 &1.602e-02 \\
$\beta$ Ari  & 0.238& 2.930e-04 &4.284e-03 &4.392e-02& 0.412 &7.343e-05 &2.118e-03 &2.473e-02 \\
$\beta$ Pic   & 0.132& 1.624e-04 &7.119e-03 &1.977e-01& 0.227 &1.215e-04 &4.499e-03 &1.951e-01 \\
$\zeta$ Lep  & 0.186& 1.429e-04 &1.937e-03 &1.677e-02& 0.326 &7.632e-05 &1.105e-03 &1.110e-02 \\
$\zeta$ Vir  & 0.163& 9.323e-05 &1.223e-03 &1.032e-02& 0.281 &6.754e-05 &7.051e-04 &8.173e-03 \\
Alphekka & 0.392& 1.116e-04 &1.716e-03 &1.669e-02& 0.686 &1.995e-05 &3.451e-04 &8.738e-03 \\
$\kappa$ Phe & 0.122& 3.232e-05 &4.153e-04 &4.736e-03& 0.21 &2.410e-05 &2.365e-04 &4.006e-03 \\
Regulus  & 0.501& 1.667e-04 &3.593e-03 &2.903e-02& 0.652 &9.234e-05 &2.217e-03 &2.262e-02 \\
Mizar& 0.218& 2.337e-04 &3.983e-03 &3.775e-02& 0.378 &8.502e-05 &2.326e-03 &2.463e-02 \\
Merak & 0.314& 1.359e-04 &2.133e-03 &1.708e-02& 0.545 &2.456e-05 &4.805e-04 &7.362e-03 \\
Alioth & 0.37& 1.218e-04 &1.839e-03 &1.501e-02& 0.64 &2.524e-05 &4.269e-04 &6.740e-03 \\
$\delta$ UMa & 0.147& 1.105e-04 &1.807e-03 &2.979e-02& 0.257 &7.975e-05 &1.036e-03 &2.687e-02 \\
$\gamma$ Cet & 0.159& 6.349e-05 &8.366e-04 &7.971e-03& 0.275 &4.581e-05 &4.785e-04 &6.447e-03 \\
$\beta$ Aur  & 0.263& 2.716e-04 &3.904e-03 &2.989e-02& 0.457 &6.343e-05 &1.341e-03 &1.210e-02 \\
$\zeta$ Aql & 0.251& 1.077e-04 &1.851e-03 &1.785e-02& 0.437 &2.551e-05 &7.315e-04 &1.057e-02 \\
Phad    & 0.302& 1.295e-04 &2.145e-03 &1.709e-02& 0.524 &2.258e-05 &4.835e-04 &7.640e-03 \\
$\gamma$ Crt & 0.146& 2.947e-05 &3.360e-04 &5.540e-03& 0.251 &2.108e-05 &2.559e-04 &4.697e-03 \\
$\tau$ Eri & 0.121& 3.706e-05 &4.682e-04 &4.678e-03& 0.209 &2.773e-05 &2.674e-04 &3.881e-03 \\
$\delta$ Crv & 0.513& 1.717e-05 &2.698e-04 &9.126e-03& 0.938 &1.614e-06 &3.525e-05 &7.316e-03 \\
21 LMi   & 0.097& 2.062e-05 &2.434e-04 &2.870e-03& 0.168 &1.580e-05 &1.373e-04 &2.445e-03 \\
Algol  & 0.454& 2.054e-04 &3.959e-03 &4.293e-02& 0.847 &3.718e-05 &6.970e-04 &2.832e-02 \\
GJ 1109  & 0.102& 2.396e-05 &2.930e-04 &3.562e-03& 0.176 &1.826e-05 &1.659e-04 &3.054e-03 \\
$\lambda$ Gem & 0.156& 3.465e-05 &5.415e-04 &6.135e-03& 0.269 &2.519e-05 &3.134e-04 &5.283e-03 \\
$\gamma$ Oph & 0.183& 4.415e-05 &6.821e-04 &8.999e-03& 0.319 &2.461e-05 &3.892e-04 &7.213e-03 \\
$\beta$ Cir& 0.13& 2.535e-05 &3.074e-04 &6.010e-03& 0.224 &1.843e-05 &2.113e-04 &5.282e-03 \\
Alpheratz& 0.483& 1.240e-04 &3.008e-03 &4.256e-02& 0.902 &2.455e-05 &6.232e-04 &3.440e-02 \\
$\iota$ Boo & 0.09& 1.624e-05 &1.932e-04 &2.887e-03& 0.154 &1.253e-05 &1.085e-04 &2.538e-03 \\
\enddata

\end{deluxetable*}

\end{document}